\documentclass[aps,prd,onecolumn,showpacs,showkeys,superscriptaddress,preprintnumbers,floatfix,reprint]{revtex4}
\usepackage{graphicx}
\usepackage{amsmath}
\usepackage{bm}
\usepackage{yhmath}
\usepackage{hyperref}
\usepackage{subfigure}
\usepackage{color}
\usepackage{cases}
\usepackage{subfigure}
\usepackage{dcolumn,booktabs,bm}

\usepackage{amsfonts,amssymb,stmaryrd,latexsym,amsmath}
\usepackage{textcomp}

\usepackage{array}
\newcounter{RomanNumber}

\newcommand{\lyxmathsym}[1]{\ifmmode\begingroup\def\b@ld{bold}
  \text{\ifx\math@version\b@ld\bfseries\fi#1}\endgroup\else#1\fi}

\def\mud{{\lambda}}

\allowdisplaybreaks

\UseRawInputEncoding
\begin{document}

\title{Spin-$\frac{3}{2}$ doubly charmed baryon contribution to the magnetic moments of the spin-$\frac{1}{2}$ doubly charmed baryons}

\author{Hao-Song Li}\email{haosongli@nwu.edu.cn}\affiliation{Institute of Modern Physics and School of Physics, Northwest University, Xian 710127, China}\affiliation{Shaanxi Key Laboratory for Theoretical Physics Frontiers, Xi¡¯an 710127, China}\affiliation{
Peng Huanwu Center for Fundamental Theory, Xi¡¯an 710127, China}

\author{Wen-Li Yang}
\affiliation{Institute of Modern Physics and School of Physics, Northwest University, Xian 710127, China}\affiliation{Shaanxi Key Laboratory for Theoretical Physics Frontiers, Xi¡¯an 710127, China}\affiliation{
Peng Huanwu Center for Fundamental Theory, Xi¡¯an 710127, China}

\begin{abstract}
We have systematically investigated the magnetic moments of spin-$\frac{1}{2}$ doubly charmed baryons in the framework of the heavy baryon chiral perturbation theory. In this
paper, one loop corrections with intermediate spin-$\frac{1}{2}$ and spin-$\frac{3}{2}$ doubly charmed baryon states are considered. The numerical results are calculated to next-to-leading order: $\mu_{\Xi^{++}_{cc}}=0.35\mu_{N}$, $\mu_{\Xi^{+}_{cc}}=0.62\mu_{N}$, $\mu_{\Omega^{+}_{cc}}=0.41\mu_{N}$. Our results may be useful for future experiment and chiral extrapolation of the lattice QCD.
\end{abstract}

\maketitle

\thispagestyle{empty}

\section{Introduction}\label{Sec1}

The doubly charmed baryon was first claimed by SELEX Collaboration in the decay mode $\Xi^{+}_{cc}\rightarrow\Lambda_{c}^{+}K^{-}\pi^{+}$ with the mass $M_{\Xi^{+}_{cc}}=3519\pm1\rm{MeV}$~\cite{Mattson:2002vu}. Since then, many experimental collaborations have shown great interest to the doubly charmed baryons. Although other experimental collaborations like FOCUS~\cite{Ratti:2003ez}, BABAR~\cite{Aubert:2006qw} and Belle~\cite{Chistov:2006zj} did not find any evidence for doubly charmed baryons, researchers have never stopped studying on the doubly charmed baryons. In 2017, LHCb
collaboration reported the discovery of $\Xi^{++}_{cc}$ in the mass spectrum of
$\Lambda_{c}^{+}K^{-}\pi^{+}\pi^{+}$  with the mass
$M_{\Xi^{++}_{cc}}=3621.40\pm0.72\pm0.27\pm0.14 \rm{MeV}$~\cite{Aaij:2017ueg}.

In the past decade, the masses and decay properties of double charmed baryons have been studied extensively in literature \cite{Bagan:1992za,Roncaglia:1995az,Ebert:1996ec,Tong:1999qs,Itoh:2000um,Gershtein:2000nx,
Kiselev:2001fw,Narodetskii:2001bq,Mathur:2002ce,Lewis:2001iz,Ebert:2002ig,Flynn:2003vz,AliKhan:1999yb,Brambilla:2005yk,Vijande:2004at,Chiu:2005zc,
Migura:2006ep,Albertus:2006ya,Liu:2007fg,Roberts:2007ni,Valcarce:2008dr,Liu:2009jc,Namekawa:2012mp,Alexandrou:2012xk,
Aliev:2012ru,Aliev:2012iv,Namekawa:2013vu,Sun:2014aya,Chen:2015kpa,Sun:2016wzh,
Shah:2016vmd,Chen:2016spr,Kiselev:2017eic,Chen:2017sbg,Yao:2018ifh,Ozdem:2018uue,Bahtiyar:2018vub,Meng:2018zbl,Wang:2019mhm}.
It is very important to investigate the baryon electromagnetic form factors, especially the magnetic moments as the electromagnetic properties give information about the internal structures and shape deformations, which provide valuable
insight in describing the inner structures of hadrons and understanding the mechanism of strong interactions at low-energy.
With nonrelativistic qurak model, Lichtenberg first investigated the doubly charmed baryons magnetic moments in Ref.~\cite{Lichtenberg:1976fi}. With relativistic quark model, the magnetic moments have also been evaluated in Refs.~\cite{Faessler:2006ft,JuliaDiaz:2004vh}. Besides the quark models, magnetic moments have been studied in different theoretical
models and approaches \cite{Bose:1980vy,Bernotas:2012nz,Jena:1986xs,Oh:1991ws,Patel:2008xs,Can:2013zpa,Can:2013tna}, however it is difficult to include the chiral corrections. In fact, Chiral perturbation theory (ChPT)~\cite{Weinberg:1978kz} and heavy baryon chiral perturbation theory (HBChPT)~\cite{Jenkins:1990jv,Jenkins:1992pi,Bernard:1992qa,Bernard:1995dp} are quite helpful to analyze the low-energy interactions order by order. To
consider the chiral corrections, the magnetic moments of spin-$\frac{1}{2}$ doubly charmed baryons have been investigated with HBChPT in Ref.~\cite{Li:2017cfz}. In Ref.~\cite{Blin:2018pmj}, the electromagnetic form factors of the doubly charmed baryons have been studied in covariant chiral perturbation
theory within the extended on-mass-shell (EOMS) scheme. However, the spin-$\frac{3}{2}$ doubly charmed baryon was
not included explicitly in the analysis of Refs~\cite{Li:2017cfz,Blin:2018pmj}.

In this work, we will investigate the magnetic moments of spin-$\frac{1}{2}$ doubly charmed baryons in HBChPT. We explicitly consider both the spin-$\frac{3}{2}$ and spin-$\frac{1}{2}$ doubly charmed baryon intermediate states in the loop calculation because the
mass splitting between the octet and decuplet baryons is
small. Moreover, the spin-$\frac{3}{2}$ doubly charmed baryon generally strongly
couple to the spin-$\frac{1}{2}$ doubly charmed baryon. We use quark model to determine the corresponding low energy constants (LECs) and calculate the magnetic moments order by order. The analytical and numerical results are calculated to next-to-leading order.

This paper is organized as follows. In Section \ref{Sec3}, we
discuss the electromagnetic form
factors of spin-$\frac{1}{2}$ doubly charmed baryons. We introduce the effective chiral Lagrangians of spin-$\frac{3}{2}$ and spin-$\frac{1}{2}$ doubly charmed baryons in Section \ref{Sec2}. We calculate the
magnetic moments of the spin-$\frac{1}{2}$ doubly charmed baryons order by order in Section
\ref{secFormalism} and present our numerical results in Section \ref{Sec6}.
A short summary is given in Section \ref{Sec7}. We collect some useful formulae in the
Appendix~\ref{appendix-A}.

\section{Electromagnetic form factors}\label{Sec3}

For spin-$\frac{1}{2}$ doubly charmed baryons, the matrix elements of the electromagnetic current read,
\begin{equation}
<\Psi(p^{\prime})|J_{\mu}|\Psi(p)>=e\bar{u}(p^{\prime})\mathcal{O}_{\mu}(p^{\prime},p)u(p),
\end{equation}
with
\begin{equation}
\mathcal{O}_{\mu}(p^{\prime},p)=\frac{1}{M_H}[P_{\mu}G_{E}(q^{2})+\frac{i\sigma_{\mu\nu}q^{\nu}}{2}G_{M}(q^{2})].
\label{eq_new_current}
\end{equation}
where $P=\frac{1}{2}(p^{\prime}+p)$,
$q=p^{\prime}-p$, $M_{H}$ is doubly charmed baryon mass.

In the heavy baryon limit, the doubly charmed baryon field $B$ can be decomposed
into the large component $H$ and the small component
$L$.
\begin{equation}
B=e^{-iM_{H}v\cdot x}(H+L),
\end{equation}
\begin{equation}
H=e^{iM_{H}v\cdot x}\frac{1+v\hspace{-0.5em}/}{2}B,~
L=e^{iM_{H}v\cdot x}\frac{1-v\hspace{-0.5em}/}{2}B,
\end{equation}
where $M_{H}$ is the spin-$\frac{1}{2}$ doubly charmed baryon mass and $v_{\mu}=(1,\vec{0})$ is the velocity of the baryon. Now the doubly charmed baryon matrix elements of
the electromagnetic current $J_{\mu}$ can be parametrized as
\begin{equation}
<H(p^{\prime})|J_{\mu}|H(p)>=e\bar{u}(p^{\prime})\mathcal{O}_{\mu}(p^{\prime},p)u(p)\label{eq:ocurrent}.
\end{equation}
The tensor $\mathcal{O}_{\mu}$ can be parameterized in terms of
electric and magnetic form
factors.
\begin{eqnarray}
\mathcal{O}_{\mu}(p^{\prime},p)=v_{\mu}G_{E}(q^{2})+\frac{[S^{\mu},S^{\nu}]q^{\nu}}{M}G_{M}(q^{2}), \label{eq_newnew_current}
\end{eqnarray}
where $G_{E}(q^{2})$ is the electric form
factor and $G_{M}(q^{2})$ is the magnetic form
factor. When $q^2=0$, we obtain the charge ($Q$) and magnetic moment ($\mu_{H}$),
\begin{eqnarray}
Q=G_{E}(0),
\mu_{H}=\frac{e}{2M_H}G_{M}(0). \label{eq_magneticcurrent}
\end{eqnarray}
\section{Chiral Lagrangians}\label{Sec2}

\subsection{The strong interaction chiral Lagrangians}

To calculate the chiral corrections to the magnetic moment of doubly charmed baryon, we follow the
 basic definitions of the pseudoscalar mesons and the spin-$\frac{1}{2}$ baryon chiral effective Lagrangians in Refs.~\cite{Bernard:1995dp,Li2} to construct the relevant chiral Lagrangians.
The pseudoscalar meson fields read
\begin{equation}
\phi=\left(\begin{array}{ccc}
\pi^{0}+\frac{1}{\sqrt{3}}\eta & \sqrt{2}\pi^{+} & \sqrt{2}K^{+}\\
\sqrt{2}\pi^{-} & -\pi^{0}+\frac{1}{\sqrt{3}}\eta & \sqrt{2}K^{0}\\
\sqrt{2}K^{-} & \sqrt{2}\bar{K}^{0} & -\frac{2}{\sqrt{3}}\eta
\end{array}\right).
\end{equation}
 The chiral connection and axial vector
field read\cite{Bernard:1995dp}:
\begin{equation}
\Gamma_{\mu}=\frac{1}{2}\left[u^{\dagger}(\partial_{\mu}-ir_{\mu})u+u(\partial_{\mu}-il_{\mu})u^{\dagger}\right],
\end{equation}
\begin{equation}
u_{\mu}\equiv\frac{1}{2}i\left[u^{\dagger}(\partial_{\mu}-ir_{\mu})u-u(\partial_{\mu}-il_{\mu})u^{\dagger}\right],
\end{equation}
where
\begin{equation}
u^{2}=\mathit{U}=\exp(i\phi/f_{0}).
\end{equation}
We use the pseudoscalar meson decay constants
$f_{\pi}\approx$ 92.4 MeV, $f_{K}\approx$ 113 MeV and
$f_{\eta}\approx$ 116 MeV.

The doubly charmed baryon fields read.
\begin{equation}
\Psi=\left(\begin{array}{c}
\Xi_{cc}^{*++}\\
\Xi_{cc}^{*+}\\
\Omega_{cc}^{*+}
\end{array}\right),
\Psi^{*\mu}=\left(\begin{array}{c}
\Xi_{cc}^{*++}\\
\Xi_{cc}^{*+}\\
\Omega_{cc}^{*+}
\end{array}\right)^\mu\Rightarrow\left(\begin{array}{c}
ccu\\
ccd\\
ccs
\end{array}\right)^\mu,
\end{equation}
where $\Psi$ and $\Psi^{*\mu}$ are spin-$\frac{1}{2}$  and spin-$\frac{3}{2}$  doubly charmed
baryon fields, respectively.
The leading order pseudoscalar meson and baryon interaction
Lagrangians read
\begin{eqnarray}
\hat{\mathcal{L}}_{0}^{(1)}&=&\bar{\Psi}(iD\hspace{-0.6em}/-M_{H})\Psi \nonumber\\
&&+{\bar
\Psi^{*\mu}[-g_{\mu\nu}(iD\hspace{-0.6em}/-M_{T})+i(\gamma_{\mu}
D_{\mu}+\gamma_{\nu}D_{\mu})-\gamma_{\mu}(iD\hspace{-0.6em}/+M_{T})\gamma_{\nu}]\Psi^{*\nu}},
\label{Eq:baryon01}\\
\hat{\mathcal{L}}_{\rm int}^{(1)}&=&\frac{\tilde{g}_{A}}{2}\bar{\Psi}u\hspace{-0.5em}/\gamma_{5}\Psi+\frac{\tilde{g}_{C}}{2}[\bar
\Psi^{*\mu}u_{\mu}\Psi+\bar \Psi u_{\mu}\Psi^{*\mu}]
,\label{Eq:baryon02}
\end{eqnarray}
where $M_{H}$ is the spin-$\frac{1}{2}$ doubly charmed baryon mass, $M_{T}$ is the spin-$\frac{3}{2}$ doubly charmed baryon mass,
\begin{eqnarray}
D_{\mu}\Psi&=&\partial_{\mu}\Psi+\Gamma_{\mu}\Psi, \nonumber\\
D^{\nu}\Psi^{*\mu}&=&\partial^{\nu}\Psi^{*\mu}+\Gamma^{\nu}\Psi^{*\mu}.
\end{eqnarray}

In the framework of HBChPT, we denote the large component of the spin-$\frac{3}{2}$ doubly charmed baryon
as $T_{\mu}$. The leading order nonrelativistic
pseudoscalar meson and baryon Lagrangians read
\begin{equation}
\mathcal{L}_{0}^{(1)}=\bar{H}(iv\cdot
D)H-i\bar{T}^{\mu}(v\cdot
D-\delta)T_{\mu}, \label{Eq:baryon1}
\end{equation}
\begin{equation}
\mathcal{L}_{\rm
int}^{(1)}=\tilde{g}_{A}\bar{H}S_\mu u^\mu H+\frac{\tilde{g}_{C}}{2}[\bar{T}^{\mu}u_{\mu}H+\bar H u_{\mu}T^{\mu}],
\label{Eq:baryon2}
\end{equation}

where $S_{\mu}$ is the
covariant spin-operator, $\delta=M_{T}-M_{H}$ is the spin-$\frac{1}{2}$ and
spin-$\frac{3}{2}$ doubly charmed baryon mass splitting. The $\phi H H$
coupling $\tilde{g}_{A}$ and $\phi T H$
coupling $\tilde{g}_{C}$ can be estimated in the quark model in Section \ref{Sec6}. For the pseudoscalar
mesons  masses, we use $m_{\pi}=0.140$ GeV, $m_{K}=0.494$ GeV, and
$m_{\eta}=0.550$ GeV. We
use the nucleon masses $M_B=0.938$ GeV and the spin-$\frac{1}{2}$ doubly charmed baryon mass $M_{H}=3.62$ GeV.

%
%


\subsection{The electromagnetic chiral Lagrangians at $\mathcal{O}(p^{2})$}

The lowest order $\mathcal{O}(p^{2})$ Lagrangian contributes to the
magnetic moments of the doubly charmed baryons at the tree level
\begin{equation}
\mathcal{L}_{\mu_{H}}^{(2)}=a_{1}\frac{-i}{4M_{B}}\bar{H}[S^{\mu},S^{\nu}]\hat{F}_{\mu\nu}^{+}H+a_{2}\frac{-i}{4M_{B}}\bar{H}[S^{\mu},S^{\nu}]H{\rm Tr}(F_{\mu\nu}^{+}),
\label{Eq:MM1}
\end{equation}
where the coefficients $a_{1,2}$ are the LECs.
The chirally covariant QED field strength tensor $F_{\mu\nu}^{\pm}$
is defined as
\begin{eqnarray} \nonumber
F_{\mu\nu}^{\pm} & = & u^{\dagger}F_{\mu\nu}^{R}u\pm
uF_{\mu\nu}^{L}u^{\dagger},\\
F_{\mu\nu}^{R} & = &
\partial_{\mu}r_{\nu}-\partial_{\nu}r_{\mu}-i[r_{\mu},r_{\nu}],\\
F_{\mu\nu}^{L} & = &
\partial_{\mu}l_{\nu}-\partial_{\nu}l_{\mu}-i[l_{\mu},l_{\nu}],
\end{eqnarray}
where $r_{\mu}=l_{\mu}=-eQ_HA_{\mu}$ and $Q_H=\rm{diag}(2,1,1)$. The operator
$\hat{F}_{\mu\nu}^{+}=F_{\mu\nu}^{+}-\frac{1}{3}\rm Tr(F_{\mu\nu}^{+})$ is traceless and
transforms as the adjoint representation. Recall that the direct
product $3\otimes\bar{3} = 1\oplus8$ . Therefore, there are two independent
interaction terms in the $\mathcal{O}(p^{2})$ Lagrangians for the
magnetic moments of the doubly charmed baryons.

%
%
%
%

\section{Formalism up to one-loop level}\label{secFormalism}

We follow the standard power counting scheme as in Ref \cite{Li:2017cfz}, the chiral
order $D_{\chi}$ of a given diagram is given by~\cite{Ecker:1994gg}
\begin{equation}
D_{\chi}=4N_{L}-2I_{M}-I_{B}+\sum_{n}nN_{n}, \label{Eq:Power
counting}
\end{equation}
where $N_{L}$ is the number of loops, $I_{M}$ is the number of
internal pion lines, $I_{B}$ is the number of internal baryon lines and $N_{n}$ is the number of the vertices from the $n$th
order Lagrangians.
The chiral order of
 magnetic moments $\mu_{H}$ is $(D_\chi-1)$ based
on Eq. (\ref{eq_magneticcurrent}).

\begin{figure}
\centering
\includegraphics[width=0.3\hsize]{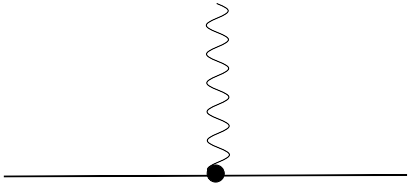}
\caption{The $\mathcal{O}(p^{2})$ tree
level diagrams where the doubly charmed baryon is denoted by the
 solid line. The dot
represent second-order coupling.}
\label{fig:tree}
\end{figure}

We assume the exact isospin symmetry with
$m_{u}=m_{d}$ throughout this work. The tree-level Lagrangians in Eq.
~(\ref{Eq:MM1}) contribute to the
doubly charmed baryon magnetic moments at $\mathcal{O}(p^{1})$ as shown in Fig.~\ref{fig:tree}. The
Clebsch-Gordan coefficients for the various doubly charmed baryons are
collected in Table~\ref{Magnetic moments}. All tree level doubly charmed baryon magnetic
moments are given in terms of $a_{1}$, $a_{2}$.

\begin{figure}[tbh]
\centering
\includegraphics[width=0.7\hsize]{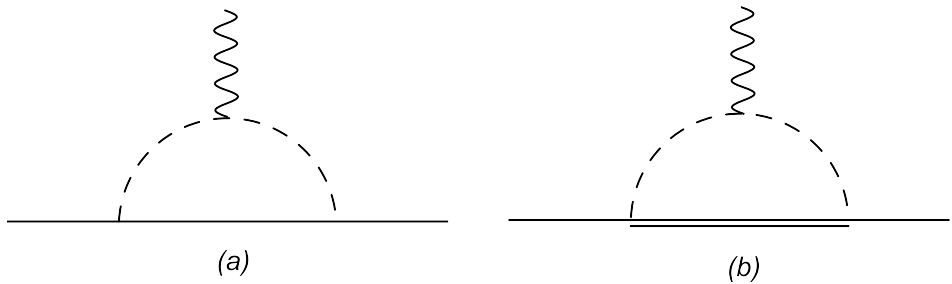}
\caption{The one-loop diagrams where spin-$\frac{1}{2}$(spin-$\frac{3}{2}$) doubly charmed baryon is
denoted by the single (double) solid line. The dashed and wiggly
lines represent the pseudoscalar meson and photon
respectively.}\label{fig:allloop}

\end{figure}

There are two Feynman diagrams contribute to the doubly charmed baryon magnetic moments to next-to-leading order at one-loop level as shown in
Fig.~\ref{fig:allloop}.
The intermediate states in diagrams (a) and (b) are spin-$\frac{1}{2}$ and spin-$\frac{3}{2}$ doubly charmed baryons respectively. The photon vertex is from the meson photon interaction term while the meson vertex is from the interaction terms in Eq.~(\ref{Eq:baryon2}).

We collect our numerical results of the doubly charmed baryon magnetic moments to the next-to-leading order in
Table~\ref{Magnetic moments}. We also compare the numerical results
of the magnetic moments when the chiral expansions are
truncated at $\mathcal{O}(p^{1})$ and $\mathcal{O}(p^{2})$ respectively in Table~\ref{various orders
Magnetic moments}.

Summing the one-loop level contributions to the doubly charmed baryon magnetic moments in Fig.~\ref{fig:allloop}, considering both the spin-$\frac{3}{2}$ and spin-$\frac{1}{2}$ doubly charmed baryon intermediate states, the loop corrections to the doubly charmed baryon magnetic moments can be expressed as
\begin{eqnarray}
\mu_{H}^{(2,\rm loop)}& = &\sum_{\phi=\pi,K}\left(-\frac{\tilde{g}_{A}^2m_{\phi}M_{N}\beta_{a}^{\phi}}{64\pi f_{\phi}^2}+\frac{\beta_{b}^{\phi}\tilde{g}_{C}^2 \left(-\delta[\log (\frac{m_{\phi}^2}{\mud ^2})-1]+2 \sqrt{m_{\phi}^2-\delta^2} \arccos(\frac{\delta }{m_{\phi}})\right)}{192 \pi ^2 f_{\phi}^2}\right)
\label{eq:mu2Loop}
\end{eqnarray}
where $\mud=4\pi f_{\pi}$ is the renormalization scale. Here, the coefficients
$\beta^\phi_{a-b}$ arise
from the spin-$\frac{1}{2}$ and spin-$\frac{3}{2}$ doubly charmed baryon intermediate states respectively in Fig.~\ref{fig:allloop}. We
collect their explicit expressions in Tables~\ref{table:ab}.

\begin{table}
  \centering
\begin{tabular}{c|cccc}
\toprule[1pt]\toprule[1pt]
$\ \ $Baryons$\ \ $ & $\ \ \beta_{a}^{\pi}\ \ \ \ $ & $\ \ \beta_{a}^{K}\ \ $ & $\ \ \beta_{b}^{\pi}\ \ $ & $\ \ \beta_{b}^{K}\ \ $ \tabularnewline
\midrule[1pt]
$\Xi_{cc}^{++}$ & $2$ & $2$ & $2$ & $2$ \tabularnewline
\hline
$\Xi_{cc}^{+}$ & $-2$ & 0 & -2 & 0 \tabularnewline
\hline
$\Omega_{cc}^{+}$ & 0 & $-2$ & 0 & -2 \tabularnewline
\bottomrule[1pt]\bottomrule[1pt]
\end{tabular}
\caption{The coefficients of the loop corrections to the
 doubly charmed baryon magnetic moments from Figs.
\ref{fig:allloop}(a) and \ref{fig:allloop}(b).} \label{table:ab}
\end{table}

With the low energy counter terms and loop contributions
(\ref{eq:mu2Loop}), we obtain the magnetic
moments,
\begin{equation}
\mu_{H}=\left\{\mu_{H}^{(1)}\right\}+\left\{\mu_{H}^{(2,\rm
loop)}\right\}
\end{equation}
where $\mu_{H}^{(1)}$
is the tree-level magnetic moments from
Eqs.~(\ref{Eq:MM1}).

\section{NUMERICAL RESULTS AND DISCUSSIONS}\label{Sec6}
\begin{table}
  \centering
\begin{tabular}{c|ccccc}
\toprule[1pt]\toprule[1pt]
Baryons & $\mathcal{O}(p^{1})$ tree &  $\mathcal{O}(p^{2})$ loop (a) &$\mathcal{O}(p^{2})$ loop (b)  \tabularnewline
\midrule[1pt]
$\Xi_{cc}^{++}$ & $\frac{2}{3}a_{1}+4a_{2}$ & $-0.51\tilde{g}_{A}^2$ & $0.20\tilde{g}_{C}^2$ \tabularnewline
$\Xi_{cc}^{+}$ & $-\frac{1}{3}$$a_{1}+4a_{2}$ &$0.15\tilde{g}_{A}^2$ & $-0.07\tilde{g}_{C}^2$ \tabularnewline
$\Omega_{cc}^{+}$ & $-\frac{1}{3}a_{1}+4a_{2}$ & $0.36\tilde{g}_{A}^2$ & $-0.12\tilde{g}_{C}^2$  \tabularnewline
\bottomrule[1pt]\bottomrule[1pt]
\end{tabular}
\caption{The doubly charmed baryon magnetic moments to the next-to-leading order(in
unit of $\mu_{N}$).} \label{Magnetic moments}
\end{table}
As there are not any experimental data of the doubly charmed baryon magnetic moments so far, in this paper,
we adopt the same strategy as in Ref. \cite{Li:2017pxa}. We use quark model to determine the leading-order tree level magnetic moments. In other words, our calculation is actually the chiral corrections to the quark model.

We collect our numerical results of the doubly charmed baryon magnetic moments to the next-to-leading order in
Table~\ref{Magnetic moments}. We also compare the numerical results
of the magnetic moments when the chiral expansions are
truncated at $\mathcal{O}(p^{1})$ and $\mathcal{O}(p^{2})$ respectively in Table~\ref{various orders Magnetic moments}.

At the leading order $\mathcal{O}(p^{1})$, as the charge
matrix $Q_{H}$ is not traceless, there are two unknown LECs
$a_{1,2}$ in Eq.~(\ref{Eq:MM1}). Notice the second column
in Table~\ref{Magnetic moments}, the $a_1$ parts are proportional to
the light quark charge within the doubly charmed baryon and the
$a_2$ parts are the same for all three doubly charmed baryons
and arise solely from the two charm quarks.

At the quark level, the flavor and spin wave functions of the
spin-$\frac{1}{2}$ and spin-$\frac{3}{2}$ doubly charmed baryons
$\Xi_{ccq}$ and $\Xi_{ccq}^{*}$ read:
\begin{eqnarray}
|\Xi_{ccq};s_3=\frac{1}{2}\rangle&=&\frac{1}{3\sqrt{2}}[2c\uparrow
c\uparrow q\downarrow-c\uparrow c\downarrow q\uparrow-c\downarrow
c\uparrow q\uparrow +2c\uparrow q\downarrow c\uparrow-c\downarrow
q\uparrow c\uparrow\nonumber\\&&-c\downarrow q\downarrow
c\downarrow+2q\downarrow c\uparrow c\uparrow-q\downarrow c\downarrow
c\downarrow-q\uparrow c\downarrow c\uparrow],\label{xiwavefunc}\\
|\Xi_{ccq}^{*};s_3=\frac{1}{2}\rangle&=&\frac{1}{\sqrt{3}}[
c\uparrow c\uparrow q\downarrow+c\uparrow c\downarrow
q\uparrow+c\downarrow c\uparrow q\uparrow],\label{xi*wavefunc}
\end{eqnarray}
where the arrows denote the third-components of the spin. $q$
can be $u$,$d$ and $s$ quark. The magnetic moments
in the quark model are the matrix elements of the following operator
,
\begin{equation}
\vec{\mu}=\sum_i\mu_i\vec{\sigma}^i, \label{magmomen}
\end{equation}
where $\mu_i$ is the magnetic moment of the quark:
\begin{equation}
\mu_i={e_i\over 2m_i},\quad i=u,d,s,c.
\end{equation}
We adopt the $m_u=m_d=336$ MeV, $m_s=540$ MeV, $m_c=1660$ MeV as the
constituent quark masses and give the results in the second column
in Table~\ref{various orders Magnetic moments}.

Up to $\mathcal{O}(p^{2})$, we need to include both the
tree-level magnetic moments and the $\mathcal{O}(p^{2})$ loop
corrections. We use the quark model to estimate the leading-order
tree level transition magnetic moments. Thus, there exist only two
LECs: $\tilde{g}_{A}$, and $\tilde{g}_{C}$ at this
order. $\tilde{g}_{A}=-0.5$ has been estimated in
Ref. \cite{Li:2017cfz}. Similarly, we can also obtain the $\phi HT$ coupling $\tilde{g}_{C}$.

At quark level, the pion-quark interaction reads
\begin{equation}
\mathcal{L}_{\rm quark}={g_0\over 2} (\bar{u}\gamma^\mu\gamma_5\partial_\mu{\pi^0}u-\bar{d}\gamma^\mu\gamma_5\partial_\mu\pi^0 d)
\end{equation}
where $g_0$ is the quark level coupling constant. The matrix elements both at hadron level and quark level,
\begin{equation}
\langle H_1|i\mathcal{L}_{H_1H_2\pi^0}|H_2;\pi^0\rangle=\langle H_1|i\mathcal{L}_{\rm quark}|H_2;\pi^0\rangle,
\end{equation}
where $H_{1,2}$ are the hadrons.
Considering the $\pi^0$ coupling at hadron level
\begin{equation}
\mathcal{L}_{\Xi_{cc}^{*++}\Xi_{cc}^{++}\pi^0}=-\frac{1}{2F_{0}}\frac{\tilde{g}_{C}}{2}\sqrt{\frac{2}{3}}
\bar{\Xi}_{cc}^{++}\partial_{\mu}\pi^{0}\Xi_{cc}^{*++}.
\end{equation}
we obtain the $\Xi_{cc}^{*++}\Xi_{cc}^{++}\pi^0$ matrix elements at the hadron level,
\begin{equation}
\langle \Xi_{cc}^{++}|i\mathcal{L}_{\Xi_{cc}^{*++}\Xi_{cc}^{++}\pi^0}|\Xi_{cc}^{*++};\pi^0\rangle
\sim-\frac{1}{2}\frac{\tilde{g}_{C}}{2}\sqrt{\frac{2}{3}}q_3,
\end{equation}
and at the quark level,
\begin{equation}
\langle \Xi_{cc}^{++}|i\mathcal{L}_{\rm quark}|\Xi_{cc}^{*++};\pi^0\rangle
\sim-\frac{2\sqrt{2}}{6}q_3.
\end{equation}
Compare with the axial charge of the nucleon,
\begin{equation}
\frac{\frac{1}{2}\frac{\tilde{g}_{C}}{2}\sqrt{\frac{2}{3}}}{-\frac{2\sqrt{2}}{6}}
=\frac{\frac{1}{2}g_{A}}{\frac{5}{6}}.
\end{equation}
We obtain the $\phi T H$
coupling $\tilde{g}_{C}=-\frac{4\sqrt{3}}{5}g_{A}=-1.75$.

With $\tilde{g}_{A}$ and $\tilde{g}_{C}$, we also need the spin-$\frac{1}{2}$ and
spin-$\frac{3}{2}$ doubly charmed baryon mass splitting $\delta=M_{T}-M_{H}$ to obtain the numerical results of the $\mathcal{O}(p^{2})$ doubly charmed baryon magnetic moments. As the spin-$\frac{3}{2}$ doubly charmed baryons have not been observed in the experiments, the masses of the spin-$\frac{3}{2}$ doubly charmed baryons remain unknown. There are independent determinations of mass splittings in nonrelativistic lattice QCD \cite{Mathur:2002ce}, lattice QCD \cite{AliKhan:1999yb,Lewis:2001iz,Flynn:2003vz} and pNRQCD \cite{Brambilla:2005yk}, the spin-$\frac{1}{2}$ and
spin-$\frac{3}{2}$ doubly charmed baryon mass splitting $\delta$ vary from 40 MeV to 120 MeV.

\begin{table}
  \centering
\begin{tabular}{c|cccc}
\hline\toprule[1pt]\toprule[1pt] Baryons($\delta=0.1$ GeV)& $\mathcal{O}(p^{1})$& $\mathcal{O}(p^{2})$ loop & $\mathcal{O}(p^{2})$ total\tabularnewline
\midrule[1pt]
$\Xi_{cc}^{++}$ &  ${4\over3}\mu_c-{1\over3}\mu_u=-0.12$ & $0.47$ & 0.35\tabularnewline
\hline
$\Xi_{cc}^{+}$ &  ${4\over3}\mu_c-{1\over3}\mu_d=0.81$ & $-0.19$ & 0.62\tabularnewline
\hline
$\Omega_{cc}^{+}$ &  ${4\over3}\mu_c-{1\over3}\mu_s=0.69$ & $-0.28$ & 0.41\tabularnewline
\bottomrule[1pt]\bottomrule[1pt]
\end{tabular}
\caption{The doubly charmed baryon magnetic moments when the
chiral expansion is truncated at $\mathcal{O}(p^{1})$ and
$\mathcal{O}(p^{2})$, respectively (in
unit of $\mu_{N}$).} \label{various orders Magnetic moments}
\end{table}

\begin{figure}[tbh]
\centering
\includegraphics[width=0.6\hsize]{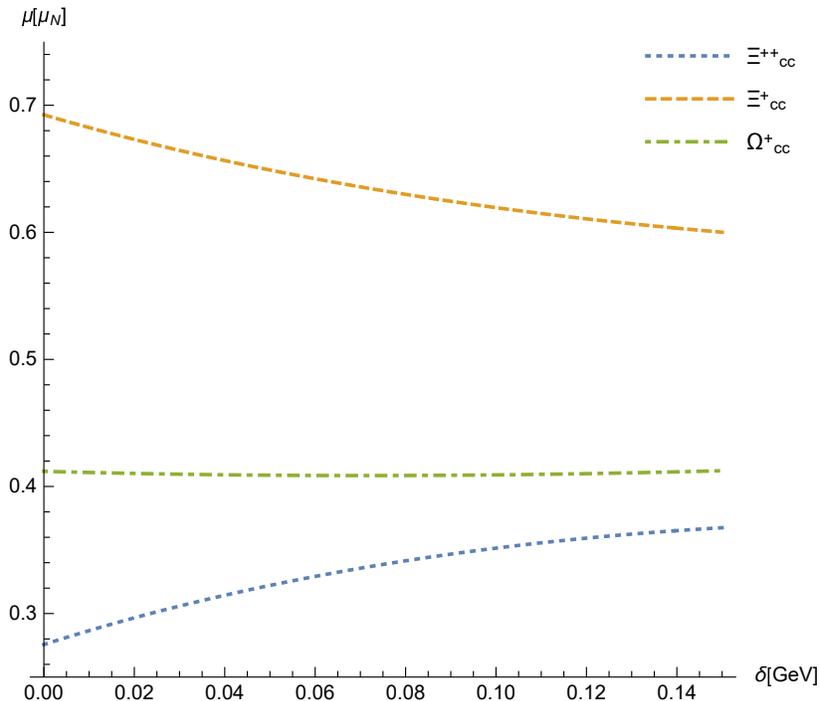}
\caption{The variations of doubly charmed baryon magnetic moments $\mu$ (in
unit of $\mu_{N}$)
with $\delta$ (in
unit of GeV).}\label{fig:pic}

\end{figure}

We adopt $\delta = 0.1$ GeV approximatively in Table~\ref{various orders Magnetic moments}, and show the variations of doubly charmed baryon magnetic moments with the spin-$\frac{1}{2}$ and
spin-$\frac{3}{2}$ doubly charmed baryon mass splitting $\delta$ in Fig.~\ref{fig:pic}. Compared to the results in Ref.~\cite{Li:2017cfz}, it is interesting to notice that after taking the spin-$\frac{3}{2}$ doubly charmed baryon contribution into consideration, the $\mathcal{O}(p^{2})$ loop corrections for the $\Xi_{cc}^{++}$ magnetic moment change from -0.13 $\mu_{N}$ to 0.47 $\mu_{N}$. The reason for this is that the $\phi T H$
coupling $\tilde{g}_{C}=-1.75$ is much bugger than the $\phi H H$
coupling $\tilde{g}_{A}=-0.5$. Thus, the spin-$\frac{3}{2}$ doubly charmed baryon contribution is magnified. The total $\Xi_{cc}^{++}$ magnetic moment changes from -0.25 $\mu_{N}$ to 0.35 $\mu_{N}$. We are looking forward to further progresses in  experiment.

 In Table~\ref{Comparison of magnetic moments}, we
compare our results obtained in the HBChPT with those from other
model calculations such as quark model (QM)~\cite{Lichtenberg:1976fi}, relativistic three-quark model (RTQM)~\cite{Faessler:2006ft}, relativistic quark
model (RQM)~\cite{JuliaDiaz:2004vh}, skyrmion description~\cite{Oh:1991ws}, confining logarithmic potential (CLP)~\cite{Jena:1986xs}, MIT bag model~\cite{Bose:1980vy}, nonrelativistic quark model (NQM)~\cite{Patel:2008xs}, lattice QCD(LQCD)\cite{Can:2013tna} and chiral
perturbation theory (ChPT)~\cite{Li:2017cfz}.

\begin{table}
  \centering
\begin{tabular}{c|ccc}
\toprule[1pt]\toprule[1pt]
Baryons & $\ \ $$\Xi_{cc}^{++}$$\ \ $ & $\ \ $$\Xi_{cc}^{+}$$\ \ $ & $\ \ $$\Omega_{cc}^{+}$$\ \ $\tabularnewline
\midrule[1pt]
QM~\cite{Lichtenberg:1976fi} & -0.124 & 0.806 & 0.688\tabularnewline
\hline
RTQM \cite{Faessler:2006ft} & 0.13 & 0.72 & 0.67\tabularnewline
\hline
RQM \cite{JuliaDiaz:2004vh} & -0.10 & 0.86 & 0.72\tabularnewline
\hline
Skyrmion \cite{Oh:1991ws} & -0.47 & 0.98 & 0.59\tabularnewline
\hline
CLP \cite{Jena:1986xs} & -0.154 & 0.778 & 0.657\tabularnewline
\hline
$\ $MIT bag model \cite{Bose:1980vy}$\ $ & 0.17 & 0.86 & 0.84\tabularnewline
\hline
NQM \cite{Patel:2008xs} & -0.208 & 0.785 & 0.635\tabularnewline
\hline
LQCD \cite{Can:2013tna} & \textemdash{} & 0.425 & 0.413\tabularnewline
\hline
ChPT
\cite{Li:2017cfz}& -0.25 & 0.85 & 0.78\tabularnewline
\hline
This work & 0.35 & 0.62 & 0.41\tabularnewline
\bottomrule[1pt]\bottomrule[1pt]
\end{tabular}
\caption{Comparison of the decuplet to octet baryon transition magnetic
moments in literature including quark model (QM)~\cite{Lichtenberg:1976fi}, relativistic three-quark model (RTQM)~\cite{Faessler:2006ft}, relativistic quark
model (RQM)~\cite{JuliaDiaz:2004vh}, skyrmion description~\cite{Oh:1991ws}, confining logarithmic potential (CLP)~\cite{Jena:1986xs}, MIT bag model~\cite{Bose:1980vy}, nonrelativistic quark model (NQM)~\cite{Patel:2008xs}, lattice QCD(LQCD)\cite{Can:2013tna} and chiral
perturbation theory (ChPT)~\cite{Li:2017cfz}(in unit of $\mu_{N}$).}
  \label{Comparison of magnetic moments}
 \end{table}

\section{Conclusions}\label{Sec7}

 In short summary, we have investigated the
magnetic moments for the spin-$\frac{1}{2}$ doubly charmed baryons to the next-to-leading order in the framework of HBChPT.
The spin-$\frac{3}{2}$ doubly charmed baryons were included as
an explicit degree of freedom and its contribution to the spin-$\frac{1}{2}$ doubly charmed baryon magnetic moments evaluated. We use quark model to determine the
leading-order magnetic moments and obtain the numerical results to next-to-leading order. Our calculation shows that $\mu_{\Xi^{++}_{cc}}=0.35\mu_{N}$, $\mu_{\Xi^{+}_{cc}}=0.62\mu_{N}$, $\mu_{\Omega^{+}_{cc}}=0.41\mu_{N}$.

Our analysis indicates that after taking the spin-$\frac{3}{2}$ doubly charmed baryon contribution into consideration, the $\Xi_{cc}^{++}$ magnetic moment changes a lot from -0.25 $\mu_{N}$ to 0.35 $\mu_{N}$. We are looking forward to further progresses and hope our results may be useful for future
experimental measurement of the doubly charmed baryons magnetic moments. Our analytical results may
be useful to the possible chiral extrapolation of the lattice simulations.

\section*{ACKNOWLEDGMENTS}

H. S. Li is very grateful to Shi-Lin Zhu and Zhan-Wei Liu for very helpful discussions. This project is supported by the National Natural
Science Foundation of China under Grants 11905171 and 12047502. This work is also supported
by the Double First-class University Construction Project of Northwest
University.

\begin{appendix}

%
%
\section{Integrals and loop functions} \label{appendix-A}

We collect some common integrals and loop functions in this
appendix.

\begin{eqnarray}
\Delta & = & i\int\frac{d^{d}l \,\mud^{4-d}}{(2\pi)^{d}}\frac{1}{l^{2}-m^{2}}= 2m^{2}(L(\mud)+\frac{1}{32\pi^{2}}\ln\frac{m^{2}}{\mud^{2}}),\\
L(\mud) & = &
\frac{\mud^{d-4}}{16\pi^{2}}[\frac{1}{d-4}-\frac{1}{2}(\ln4\pi+1+\Gamma^{\prime}(1))],
\end{eqnarray}
\begin{eqnarray}
I_{0}(q^{2})&=&i\int\frac{d^{d}l\,\mud^{4-d}}{(2\pi)^{d}}\frac{1}{(l^{2}-m^{2}+i\epsilon)
((l+q)^{2}-m^{2}+i\epsilon)},\ \ r=\sqrt{|1-4m^{2}/q^{2}|}\nonumber\\
&=&\begin{cases} \displaystyle
-\frac{1}{16\pi^{2}}(1-\ln\frac{m^{2}}{\mud^{2}}-r\ln|\frac{1+r}{1-r}|)+2L(\mud) & \left(q^{2}<0\right)\\ \displaystyle
-\frac{1}{16\pi^{2}}(1-\ln\frac{m^{2}}{\mud^{2}}-2r\, \arctan\frac{1}{r})+2L(\mud) & (0<q^{2}<4m^{2})\\ \displaystyle
-\frac{1}{16\pi^{2}}(1-\ln\frac{m^{2}}{\mud^{2}}-r\ln|\frac{1+r}{1-r}|+i\pi
r)+2L(\mud) & (q^{2}>4m^{2})
\end{cases},
\end{eqnarray}


\begin{equation}
i\int\frac{d^{d}l \,\mud^{4-d}}{(2\pi)^{d}}\frac{[1,l_{\alpha},l_{\alpha}l_{\beta}]}{(l^{2}-m^{2}+i\epsilon)(\omega+v\cdot
l+i\epsilon)}
=[J_{0}(\omega),v_{\alpha}J_{1}(\omega),g_{\alpha\beta}J_{2}(\omega)+v_{\alpha}v_{\beta}
J_3(\omega)],\ \ \omega=v\cdot r+\delta
\end{equation}
\begin{equation}
J_{0}(\omega)=\begin{cases}
\displaystyle
\frac{-\omega}{8\pi^{2}}(1-\ln\frac{m^{2}}{\mud^{2}})+\frac{\sqrt{\omega^{2}-m^{2}}}{4\pi^{2}}({\rm arccosh}\frac{\omega}{m}-i\pi)+4\omega L(\mud) & (\omega>m)\\ \displaystyle
\frac{-\omega}{8\pi^{2}}(1-\ln\frac{m^{2}}{\mud^{2}})+\frac{\sqrt{m^{2}-\omega^{2}}}{4\pi^{2}}\arccos\frac{-\omega}{m}+4\omega L(\mud) & (\omega^{2}<m^{2})\\ \displaystyle
\frac{-\omega}{8\pi^{2}}(1-\ln\frac{m^{2}}{\mud^{2}})-\frac{\sqrt{\omega^{2}-m^{2}}}{4\pi^{2}}{\rm
arccosh}\frac{-\omega}{m}+4\omega L(\mud) & (\omega<-m)
\end{cases},
\end{equation}

\begin{eqnarray}
J_{1}(\omega)&=&-\omega J_{0}(\omega)+\Delta,\\
J_2(\omega)&=&\frac{1}{d-1}[(m^2-\omega^2)J_0(\omega)+\omega\Delta],\\
J_3(\omega)&=&-\omega J_1(\omega)-J_2(\omega),
\end{eqnarray}


\begin{eqnarray}
L_{0}(\omega) & = &i\int\frac{d^{d}l\,\mud^{4-d}}{(2\pi)^{d}}\frac{1}{(l^{2}-m^{2}+i\epsilon)((l+q)^{2}-m^{2}+i\epsilon)(\omega+v\cdot
l+i\epsilon)},\ \ v\cdot q=0\nonumber\\
& = & \protect\begin{cases} \displaystyle
\frac{-1}{8\pi^{2}}\frac{1}{\sqrt{\omega^{2}-m^{2}}}({\rm arccosh}\frac{\omega}{m}-i\pi) & (\omega>m)\\ \displaystyle
\frac{1}{8\pi^{2}}\frac{1}{\sqrt{m^{2}-\omega^{2}}}\arccos\frac{-\omega}{m} & (\omega^{2}<m^{2})\protect\\ \displaystyle
\frac{1}{8\pi^{2}}\frac{1}{\sqrt{\omega^{2}-m^{2}}}{\rm
arccosh}\frac{-\omega}{m} & (\omega<-m)
\protect\end{cases}.
 \end{eqnarray}

\end{appendix}

\vfil \thispagestyle{empty}

\end{document}